# Vaccine escape in 2013-4 and the hydropathic evolution of glycoproteins of A/H3N2 viruses

J. C. Phillips

Dept. of Physics and Astronomy, Rutgers University, Piscataway, N. J., 08854

Abstract
EXCODD

More virulent strains of influenza virus subtypes H1N1 appeared widely in 2007 and H3N2 in 2011, and especially 2013-4, when the effectiveness of the H3N2 vaccine decreased nearly to zero. The amino acid differences of neuraminidase from prior less virulent strains appear to be small (<1%) when tabulated through sequence alignments and counting site identities and similarities. Here we show how analyzing fractal hydropathic forces responsible for neuraminidase globular compaction and modularity quantifies the mutational origins of increased virulence. It also predicts vaccine escape and specifies optimized targets for the 2015 H3N2 vaccine different from the WHO target. Unlike some earlier methods based on measuring hemagglutinin antigenic drift and ferret sera, which take several years, cover only a few candidate strains, and are ambiguous, the new methods are timely and can be completed, using NCBI and GISAID amino acid sequences only, in a few days.

**1. Introduction** A major problem facing the vaccination community is the lengthy time delay between the arrival of new viral strains, and determination of their suitability as vaccine targets. Because of these delays, there were an extra 6 million cases of 3-4 week H3N2 flu in America alone in both 2014 and 2015 seasons. We have invented a physico-chemical one-dimensional method for shortening this time substantially. The new method involves analyzing flu strains using fractal bioinformatic scaling. This method had previously been used successfully to analyze the evolution under vaccination pressures of neuraminidase (NA) H1N1 (1934-2013) and H3N2 (1968-2013) strains [1,2]. The fractal analysis of NA is much simpler than that for hemagglutinin (HA), and it has yielded strong one-dimensional results for the vaccine-escaping H3N2 2013-4 flu strains.  Specifically it has separated antigenic drift into two opposing components, based on migration and vaccination pressures, and shown that migration-driven vaccine escape can be recognized at an early stage.

There have been several attempts to predict evolving dominant new strains based on two-dimensional clustering of antigenicity and numbers of HA1 head sequence mutations in the receptor region 130-230 [3].  Five epitopes (A – E), involving 131 amino acid sites in the receptor region, were identified [3] More recently a computational method studied the predictability of similar methods based on combined sequence and antigenic data for vaccine selection from 2002 to 2007 [4].  A new physical method [5] used only old antigenicity data, and requires no new antigenicity data, only the amino acid sequences of circulating strains in the GenBank-GISAID data base.  A 2011 refinement, based on entropy, reduced 131 sites to 54, and also identified epitope A and B sites as the most rapidly evolving [6].

A recent entropy study, based on HA1 sequence  epitopes alone [7], actually predicted the failure of the 2014 H3N2 vaccine before the data showing its more than three-fold drop in effectiveness were made public [8].  It correlates with vaccine effectiveness with $R^2$ = 0.76, which is significantly higher than antigenic and/or ferret-derived distances for vaccine effectiveness in humans [7,9].  A comparative advantage of the physical methods, based only on either HA sequences [5-7] or HA and NA sequences (here), is their ability to respond in days, rather than years, to the need for a identifying an optimal vaccine target for responding to a virulent viral vaccine escape.

As noted in [1], the critical step in viral activity is probably the formation of oligomerized HA and NA glycosidic spikes. Vaccines protect against viruses by poisoning spike growth. Because the entropic methods of [5-7] rely only on HA1 sequences, while the fractal bioinformatic methods of [1] rely only on NA sequences, these two economical and complementary methods are equally rapid and appear to represent a major advance in dealing with the needs of timely vaccine design [8].  The basis for the entropic method is natural, and is easily understood [5,6].

The close correspondence of results obtained by these two physical methods [1,7] not only justifies their methodology, but it also enhances their predictive reliability.

**2. Results**  All the H3N2 trends [1,2] up to 2011 have occurred under the restraining influence of an effective vaccination program.  In 2013-4 new strains of H3N2 appeared that largely evaded the prepared vaccine (only 15% effective in 2014 and 2015, the lowest seen by CDC in the decade since it began tracking annual vaccine effectiveness) [7,8].  The most pressing question is, can any of these methods predict a timely target strain for vaccine development [9]? Is it possible to identify these few early vaccine-escaping strains automatically among the thousands of new sequences deposited in GenBank and GISAID annually?  If so, then we can hope to include the answer in the selected vaccine target strain at least a year earlier, and avoid another failure in the near future.  Here we benefit from not only having a clear-cut problem, but also from having high-resolution amino-acid sequencing tools like BLAST [1,2,5-7], as well as explicit inclusion of legacy antigenicity and protein shape data.

We begin by discussing recent changes in the HA1 head profile $<\Psi(j)W>$, where the hydropathicity scale $\Psi$ has been averaged over a sliding window W = 111, as before for H1N1 [1].  We take up the discussion of the evolution of H3 where it ended earlier [2], with Texas 2011.  The HA1 profiles of two new epitopic clusters based on candidate strains Switz 2013 and Nebraska 2014 are shown in Fig. 1.  Switzerland/9715293/2013 (GISAID) is the WHO 2015 target strain, a choice supported by [10], using glycosylation data.  However, using more recent sequence mutation data and epitopic clustering in the HA head, [7] recommended Nebraska/04/2014 (GenBank).

A fundamental theory of the meaning of hydropathic profiles has been discussed elsewhere [2].  Large values correspond to hydrophobic regions, small values to hydrophilic ones.  The profiles are obtained from bioinformatic scales [11,12] which have been transformed by sliding windows of length W.  In the case of HA, the window length W corresponds to the range of the longest H3 epitope in the receptor region, which smooths noisy oscillations.  It is not an adjustable parameter for H3N2, as the H1N1 value [1] is used unchanged.  The profile extrema (maxima or minima) can correspond mechanically to elastic pinning points or hinges.

As shown in Fig. 1, the HA1 profiles of two new epitopic clusters based on Switz 2013 and Nebraska 2014 candidates are qualitatively different.  The largest change in 2014 Nebraska is the appearance of a strongly hydrophobic shift in the B epitope [5] region 128-198, which continues to the C end of the HA1 head.  This contrasts strongly with the smaller hydrophilic shift of the Switz 2013, the WHO target strain.  Deem et al. have shown [7] that the dominant circulating strain for 2015-2016 will probably contain 2014 Nebraska, but not Switz 2013.  If so, it is unlikely that a vaccine based on Switz 2013, the WHO target strain, will be effective.  This conclusion is reinforced by the hydropathic reversal shown in Fig. 1.

Most treatments of HA evolution treat only HA1 mutations, and do not allow for differences in hydropathicity [3,4].  An alternative approach focusses on frequent mutations of single sites with large changes in hydropathicity, in common sequences with many copies. Results with this approach are shown in Table I.  Also shown are NA roughnesses with W = 17, which gave a good description of the evolution of H1N1.  This approach distinguished between strains with decreasing virulence under vaccination pressure, and increasing virulence due to new strains, such as swine flu.

The strain employed in the 2013 and 2014 H3N2 Northern hemisphere vaccines was Texas/50/2012; it is the first line in our summarized results in Table I.  To understand this Table fully, the reader should begin by studying [1,2,5-7].  Here are some salient points.  The present NA method is based on evolutionary changes in the globular shape of NA, which are monitored in terms of its overall hydropathic roughness $\Re(\Psi(N,17))$ at a scaling length of W = 17 amino acids, appropriate to membrane interactions, as previously used for H1N1 [1].  Unlike the entropic method used to study HA [5-7], which weights all substitutions equally, the present method emphasizes those substitutions X - Y that involve large changes in hydropathicity $\Delta\Psi = \Psi(X) - \Psi(Y)$.  This is because the globular shape is determined by competing hydrophobic forces (pushing segments inwards towards the globular core) and hydrophilic forces (pushing segments outwards towards the globular surface).  We do not need to know the details of the globular fold, because this is largely retained in proteins with amino acid similarities greater than 40%, and here the evolutionary changes are usually of order 1%.  To treat the hydropathic interactions most accurately, we employ the MZ fractal hydropathicity scale $\Psi$ discovered by [11]. In the present case, quite similar results are obtained from the standard KD $\Psi$ scale based on enthalpy transfer from water to air (not shown) [12].

The NA globular roughness $\Re(17)$ is the simplest factor affecting oligomer kinetics; it is listed in the last column. Between 1968 and 2011, $\Re(17)$ for N2 decreased from 210 to 165, as a result of pressure from an effective vaccine, parallel to the N1 trend [1,2]. In some cases, it was reduced even further, to 159, as for Texas/50/2012. The first sign of $H_3N_2$ vaccine escape occurred in Boston in the Fall of 2012. The large number of copies of this $\Re(17) = 167$ strain is indicative of concern among clinicians that a new strain, controlled less effectively by the 2011 vaccine, could have been emerging. Indeed, between 2009 and 2011, vaccine effectiveness decreased from ~ 50% to ~ 15% [7], which is explained by the appearance of a new "swine" strain marked by an increasing $\Re(17)$.

It had been suggested optimistically in 2008, when the H3N2 vaccine was still effective, that "the antigenic characteristics of A (H3N2) viruses outside E- SE Asia may be forecast each year based on surveillance within E- SE Asia, with consequent improvements to vaccine strain selection" [13]. However, comparison of the Singapore H2013.910 strain, $\Re(17)= 171$ (see also Fig. 2), with Cal./58/2013, $\Re(17) = 172$, shows that once vaccine escape has occurred, the new strain can occur anywhere. Similarly, study of Fig. 2 shows that although the Singapore 2013 strain has escaped, so have several other 2013 strains. Finally, note that although there is good overall agreement between the HA1 entropic cluster sequence analysis of [5-7] and the present results, [5-7] thought Cal.02/2014 dangerous, but it seems to be less dangerous than Cal./58/2013 (Table I).

The overall conclusion from Fig. 2 is that vaccine escape occurred after 2012 for all strains above $\Re(17)$ ~ 169, when the roughness trend reversed from decreasing to increasing. We agree with [7] that Nebraska 04/2014 is exceptionally virulent and has probably escaped the Texas/50/2012 vaccine almost completely. Whether one should have used virulent and wide-spread New York/04/2014 or equally virulent Nebraska 04/2014 as the target strain for the 2015 H3N2 vaccine is hard to say. Given the differences listed in Table I, a good vaccine could also be based on both Cal./58/2013 and New York/04/2014. There is one feature of this Table that is surprising. Sites 158 (A) and 176 (B) lie in the 130-230 receptor region studied by [3] and [7], but allosteric 327 does not. These 158 and 327 large $\Delta\Psi$ mutations occur in tandem in New York/04/2014 and Nebraska 04/2014, but not in earlier HA strains. It appears that the HA1 327 allosteric mutation often plays a key role in vaccine escape.

## 3. Discussion

Is it an accident that H3N2 has escaped its vaccine, whereas the H1N1 vaccines easily contained, and continue to contain, the 2007 swine flu [1]? Hemagglutinin consists of a stalk (HA2, 344-566) and head (HA1,17-344) structure. The secondary structures are primarily helical (HA2), hard, and strand (HA1), soft [14]. The hard/soft alternation provides both flexibility (important for braided oligomerization) and strength (needed for membrane penetration). Similarly, the absence of secondary structure in H3N2 NA between 316 and 353, with a turn at 331, provides flexibility relative to the beta strand structure. Overall, NA is softer, and HA harder. Thus the hard/soft alternation occurs on at least three levels, a noteworthy example of self-similar scaling as a factor promoting oligomerization to form 13 nm HA and 14 nm NA spikes [15].

Most proteins contain hinges, and the hinges involved in open-closed conformational transitions are easily identified in local density elastic network models [16]. It would be interesting to see if these models, which also identify soft vibrational modes, could yield further insights into the NA hinges identified in Fig. 3. We can use hinges to answer a question of immediate biomedical interest: is it an accident that H3N2 has escaped its vaccine, whereas the H1N1 vaccines easily contained, and continue to contain, the 2007 swine flu [1]? If we compare the N1 hydroprofile of $<\Psi(j)17>$ (Fig. 5 of [1]) with the corresponding profile of N2 (here, Fig. 3), we notice a striking difference. There is one wide and very soft hinge near 335 in N2, but there are three narrow and soft N1 hinges, near 150, 220 and 390. This means that a single softening mutation near 335 can destabilize N2, whereas multiple softening mutations would have to occur near 150, 220 and 390 simultaneously to promote N1 vaccine escape.

We noted above that the deep hydrophilic minimum of N2 near 335 is associated with a wide disordered 37 amino acid region. The three deepest hydrophilic minima of N1 are associated with shorter disorder: (150), 17; (220), 15; and (390) 13 – less than half as large disordered regions (PDB 4B7N). Softening of the central 220 N1 minimum by an I223R mutation increases antibody resistance [17] and could reduce vaccine effectiveness.

## 4. Conclusions

It is becoming increasingly clear that many aspects of compacted networks, including both glasses and proteins, are universal [18,19]. This universality across inorganic, organic, and living systems supports the apparent stability of the HA and NA bioinformatic sequencing methods for determining vaccine target strains discussed here. One would expect that application of these methods to both HA and NA sequenced stages of vaccine formation from deactivated viruses might be useful both for improving vaccine protocols and monitoring vaccine quality on-line during production, but this topic lies outside the scope of this paper.

**Methods** The calculations described here are very simple, and are most easily done on an EXCEL macro. The one used in this paper was built by Niels Voohoeve and refined by Douglass C. Allan. I have benefitted from conversations with M. W. Deem, A. Lee, and D. Spiro.

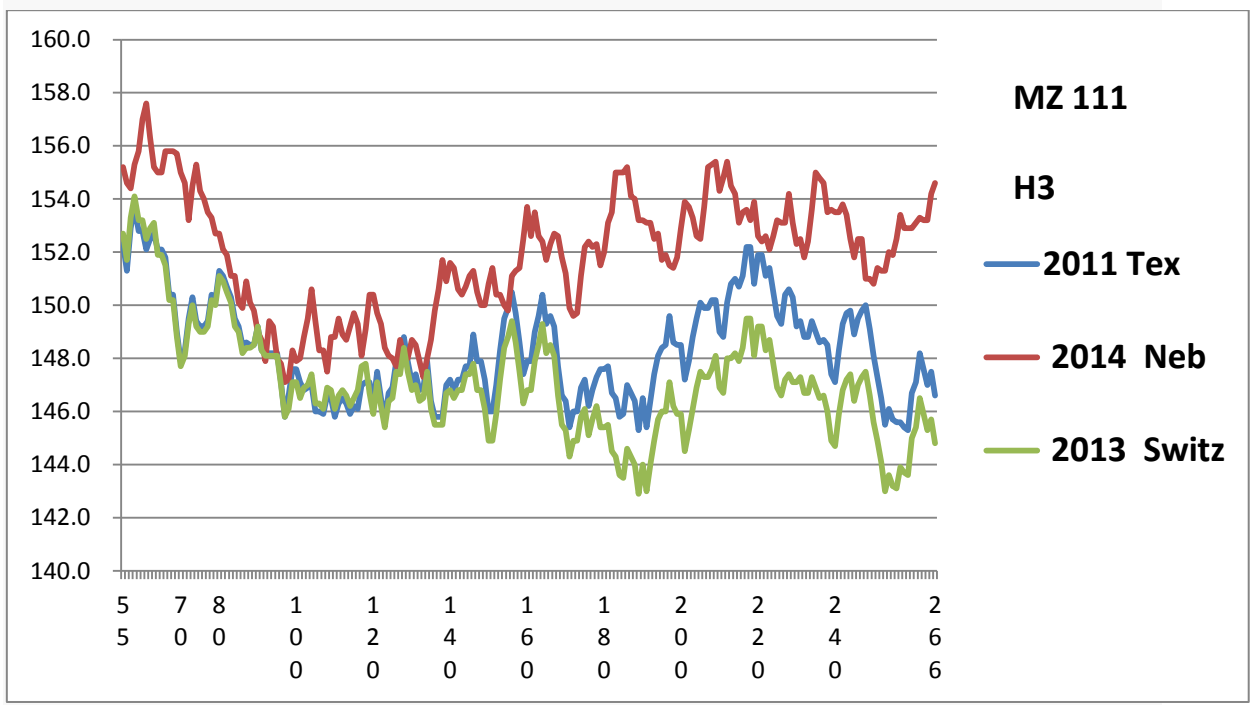

Fig. 1. The HA1 profiles of Switz 2013, the WHO target strain for 2015-2016, is similar to 2011 Texas, but slightly less hydrophobic. Two H3N2 clusters emerged in 2014 according to epitopic analysis [5-7], one represented by Switz 2013, and the other by Nebraska 2014. Note that relative to Tex 2011, the two new clusters have shifted in opposite directions, with Nebraska 2014 being much the most hydrophobic.

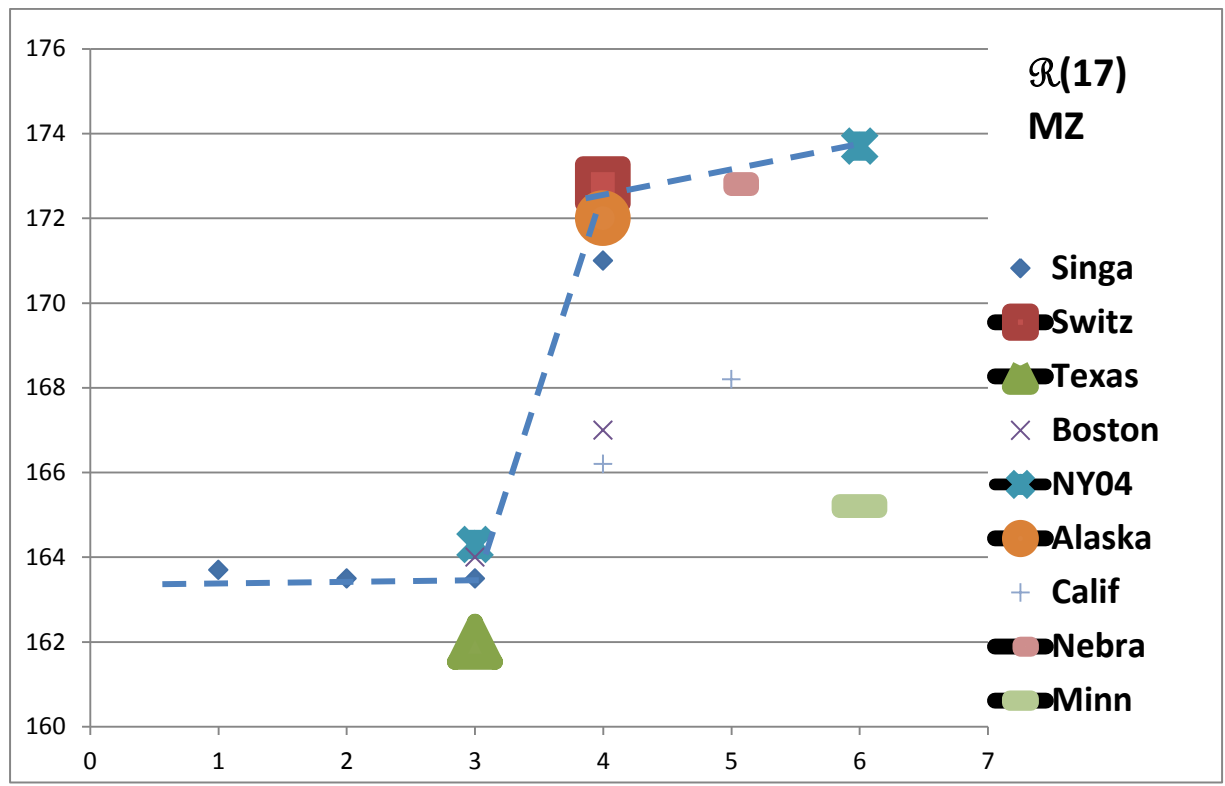

Fig. 2. A sketch of the evolution of N2 (NA) $\mathcal{R}(17)$ of H3N2 from 2010 [abscissa 1], through 2011[2], 2012 [3], 2013[4], 2014 (Jan.)[5] and 2014 Nov.[6]. The [1] - [4] specimens were chosen using BLAST to identify strains evolving towards the escaped Alaska and New York/04/2014 strains [6]. Singapore H2013.910 was collected 11/28/2013, and submitted to the NIAID - NCBI data base 6/26/2014, which was too late for use as a target strain for the 2014. The so-called "Alaska 2013" strain includes Cal./58/2013 (marked here as Alaska) and Iowa/08/2013, both collected before Dec. 15 2013. The blue dashed line follows a small number of increasingly virulent strains, increasing abruptly in 2013. Strains above 169 will escape vaccines based on Texas 2012 (large green triangle). The New York 2014 Fall strain is closely related to the Nebraska 2014 Spring strain (orange rectangle). The overall sequential non-positive mutational differences between these strains are very small, for example, only 0.7% differences between Texas 2012 and Nebraska 2014.

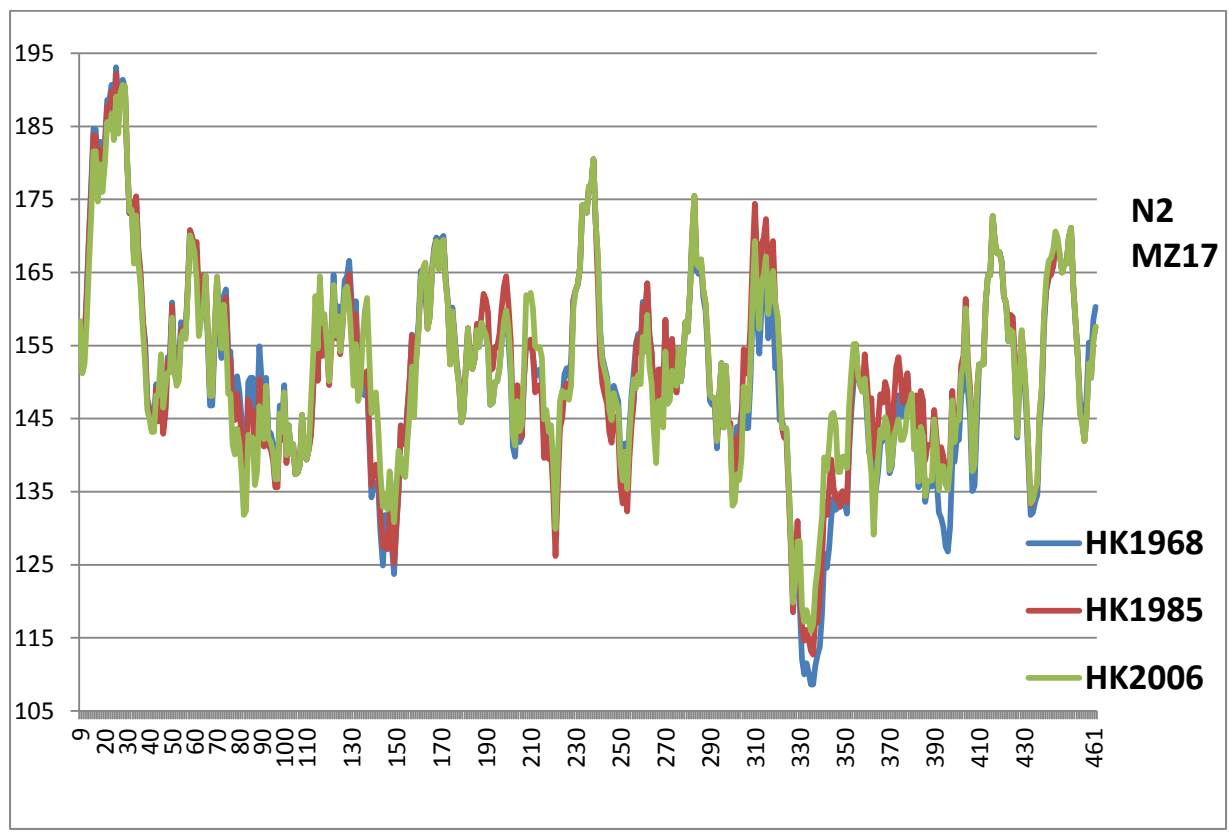

Fig. 3. The evolution of Hong Kong strains of NA from H3N2 from 1968 until 2006. Note that the largest changes occur in the two mechanical hinges (hydrophilic extrema) near 335 and 405, and correspond to increasing these minima, which decreases the overall oscillation amplitude and reduces the roughness $\Re(17)$.

| Strain | 158 | 176 | 327 | #HA Copies | NA $\Re(17)$ |
|---|---|---|---|---|---|
| Texas/50/2012 | R | K | Q | 2 | 159.3 |
| Sing. 2013 | R | K | Q | 141 | 166.6 |
| Boston YGA/09/2013 | R | K | Q | 100 | 167.0 |
| Cal./58/2013 | G | K | Q | 52 | 172.0 |
| Cal.02/2014 | G | K | Q | 75 | 168.4 |
| New York/04/2014 | R | T | H | 113 | 173.7 |
| Nebraska/04/2014* | R | T | H | 5 | 172.8 |
| Switz/2013 | G | K | Q | 12 | 172.8 |

Table I. Evolution of three HA sites and NA roughness during 2013-2014. The increase in the numbers of HA copies of the same strain in 2012-2013 seasons indicates growing clinical concern over H3N2 mutations, for example, in Boston 2012 and Singapore 2013. Two different HA mutations, 158G, and the double 176T and 327H, could have contributed 2014 strains that were successful in escaping the H3N2 2012 vaccine. Note that the allometric interactions between 176 and 327 involve shifts $\Delta\Psi$ which have the same sign and similar magnitudes: $\Delta\Psi(K176T) = 66$, and $\Delta\Psi(Q327H) = 47$, while $\Delta\Psi(R158G) = 78$. In each case the NA roughening increase is accompanied in tandem by increased HA hydrophobicity. The * strain is the target recommended for the 2015 H3N2 vaccine in [7] and here. Escape from the vaccine based on Texas/50/2012 came to be substantial above NA $\Re(17) \sim 169$, and may have already begun in Boston 2012. The WHO target for the 2015 vaccine is Switz/2013; it is discussed at the end of the paper. [7] identifies the dominant swine strain as Kansas2013. The value of NA $\Re(17)$ for this strain is 196.7. It is reasonable to describe the most virulent H3N2 human strains as "swine flu". Switz/2013 at 172.8 would appear to be a good choice for the target strain of the H3N2 2015 vaccine strain, but Nebraska 2014 is equally good, so far as NA is concerned. However, Fig. 1 for HA shows that these two strains are qualitatively different, and [7] predicted that Nebraska 2014 would be a much better target strain than the WHO choice (also [10]) Switz/2013.